\documentclass[11pt,a4paper]{article}

\usepackage{amsmath,amsfonts,amssymb,bm}

\usepackage{graphicx} 

\usepackage[hang,normalsize,bf]{subfigure} 

\begin{document}
\title{Unintegrated parton distributions \\ and particle production in
  hadronic collisions%
\thanks{Presented at the XXXIII International Symposium on
Multiparticle Dynamics, 5-11 September 2003, Cracow, Poland}%
}
\author{Antoni Szczurek \\
         Institute of Nuclear Physics\\
         PL-31-342 Cracow, Poland \\  
         Rzesz\'ow University \\
         PL-35-959 Rzesz\'ow, Poland}

\maketitle
\begin{abstract}
The inclusive distributions of gluons and pions
for high-energy NN collisions are calculated.
The results for several unintegrated gluon distributions (UGD's)
from the literature are compared.  
We find huge differences in both rapidity and $p_t$
of gluons and $\pi$'s in NN collisions
for different models of UGD's.
The Karzeev-Levin UGD gives good description
of momentum distribution of charged hadrons at midrapidities.
We find, however, that the gluonic mechanism discussed does not describe
the inclusive spectra of charged particles in the fragmentation
region. Some of the missing mechanisms are calculated with
the help of unintegrated parton distributions from
the solution of the CCFM equation.
\end{abstract}

  
\section{Introduction}
The recent results from RHIC (see e.g. \cite{RHIC}) have attracted
renewed interest in better understanding the dynamics of
particle production, not only in nuclear collisions.
Quite different approaches have been used
to describe the particle spectra from the nuclear collisions \cite{PHOBOS}.
The model in Ref.\cite{KL01} with an educated guess
for UGD describes surprisingly well
the whole charged particle rapidity distribution by means
of gluonic mechanisms only. Such a gluonic mechanism would lead to
the identical production of positively and negatively charged hadrons.
The recent results of the BRAHMS experiment \cite{BRAHMS} put into
question the successful description of Ref.\cite{KL01}.
In the light of this experiment, it becomes obvious that
the large rapidity regions have more complicated flavour structure.

I discuss the relation between UGD's in hadrons and
the inclusive momentum distribution
of particles produced in hadronic collisions.
The results obtained with different UGD's
\cite{KL01,AKMS94,GBW_glue,KMR,Blue95} are shown and compared.
I present first results obtained with unintegrated parton
distributions from the solution of the so-called CCFM equation.

\section{Inclusive gluon production}

At sufficiently high energy the cross section for inclusive
gluon production in $h_1 + h_2 \rightarrow g$ can be written
in terms of the UGD's ``in'' both colliding
hadrons \cite{GLR81}
\begin{equation}
\frac{d \sigma}{dy d^2 p_t} = \frac{16  N_c}{N_c^2 - 1}
\frac{1}{p_t^2}
\int
 \alpha_s(\Omega^2)
 {\cal F}_1(x_1,\kappa_1^2) {\cal F}_2(x_2,\kappa_2^2)
\delta(\vec{\kappa}_1+\vec{\kappa}_2 - \vec{p}_t)
\; d^2 \kappa_1 d^2 \kappa_2    \; .
\label{inclusive_glue0}
\end{equation}
Above ${\cal F}_1$ and ${\cal F}_2$ are UGD's
in hadron $h_1$ and $h_2$, respectively.
The longitudinal momentum fractions are fixed by kinematics:
$x_{1/2} = \frac{p_t}{\sqrt{s}} \cdot \exp(\pm y)$.
The argument of the running coupling constant is taken as
$\Omega^2 = \max(\kappa_1^2,\kappa_2^2,p_t^2)$.

\begin{figure}[!thb]
\begin{center}
  \begin{center}
\includegraphics[width=6cm]{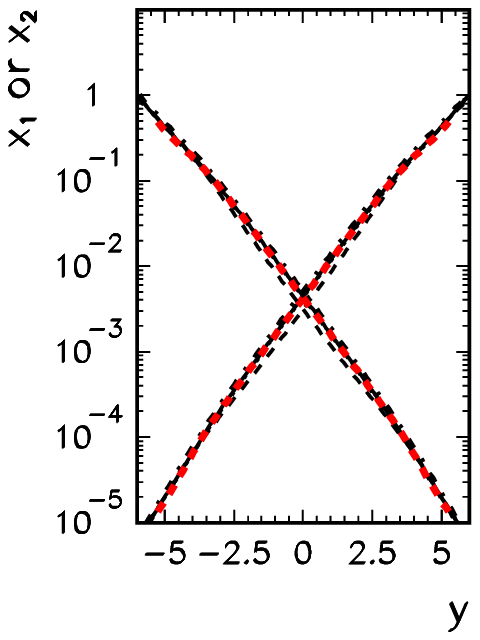}
  \end{center}
\caption[*]{
 $<x_1>$ and $<x_2>$
for $p_t >$ 0.5 GeV and at W = 200 GeV. 
\label{fig:x1_x2}
}
\end{center}
\end{figure}

Here I shall not discuss the distributions of ``produced'' gluons,
which can be found in \cite{Sz03}.
Instead I shall discuss what are typical values of $x_1$ and $x_2$
in the jet (particle) production.
Average value $<x_1>$ and $<x_2>$, shown in Fig.\ref{fig:x1_x2},
only weakly depend on the model of UGD.
For $y \sim$ 0 at the RHIC energy W = 200 GeV one tests
UGD's at $x_g$ = 10$^{-3}$ - 10$^{-2}$.
When $|y|$ grows one tests more and more asymmetric (in $x_1$ and $x_2$)
configurations. For large $|y|$ either $x_1$ is extremely
small ($x_1 <$ 10$^{-4}$) and $x_2 \rightarrow$ 1
or $x_1 \rightarrow$ 1 and $x_2$ is extremely small ($x_2 <$ 10$^{-4}$).
These are regions of gluon momentum fraction where the UGD's
is rather poorly known. The approximation used in obtaining
UGD's are valid certainly only for $x <$ 0.1.
In order to extrapolate the gluon distribution to
$x_g \rightarrow$ 1 I multiply
the gluon distributions from the previous section by a factor
$(1-x_g)^n$, where n = 5-7.


\section{From gluon to particle distributions}

In Ref.\cite{KL01} it was assumed, based on the concept
of local parton-hadron duality, that the rapidity distribution
of particles is identical to the rapidity distribution of gluons.
In the present approach I follow a different approach
which makes use of phenomenological fragmentation functions (FF's).
For our present exploratory study it seems
sufficient to assume $\theta_h = \theta_g$.
This is equivalent to $\eta_h = \eta_g = y_g$, where $\eta_h$ and
$\eta_g$ are hadron and gluon pseudorapitity, respectively. Then
\begin{equation}
y_g = \mathrm{arsinh} \left( \frac{m_{t,h}}{p_{t,h}} \sinh y_h \right)
\; ,
\label{yg_yh}
\end{equation}
where the transverse mass $m_{t,h} = \sqrt{m_h^2 + p_{t,h}^2}$.
In order to introduce phenomenological FF's
one has to define a new kinematical variable.
In accord with $e^+e^-$ and $e p$ collisions I define a 
quantity $z$ by the equation $E_h = z E_g$.
This leads to the relation
\begin{equation}
p_{t,g} = \frac{p_{t,h}}{z} J(m_{t,h},y_h) \; ,
\label{ptg_pth}
\end{equation}
where $J(m_{t,h},y_h)$ is given in Ref.\cite{Sz03}.
Now we can write the single particle distribution
in terms of the gluon distribution as follows
\begin{eqnarray}
\frac{d \sigma (\eta_h, p_{t,h})}{d \eta_h d^2 p_{t,h}} =
\int d y_g d^2 p_{t,g} \int 
dz \; D_{g \rightarrow h}(z,\mu_D^2) \\ \nonumber
\delta(y_g - \eta_h) \; 
\delta^2\left(\vec{p}_{t,h} - \frac{z \vec{p}_{t,g}}{J}\right)
\cdot \frac{d \sigma (y_g, p_{t,g})}{d y_g d^2 p_{t,g}} \; .
\label{from_gluons_to_particles}
\end{eqnarray}

In the present calculation I shall use only LO
FF's from \cite{BKK95} or \cite{Kretzer2000}.

\begin{figure}[!thb]
\begin{center}
  \begin{center}
\includegraphics[width=6cm]{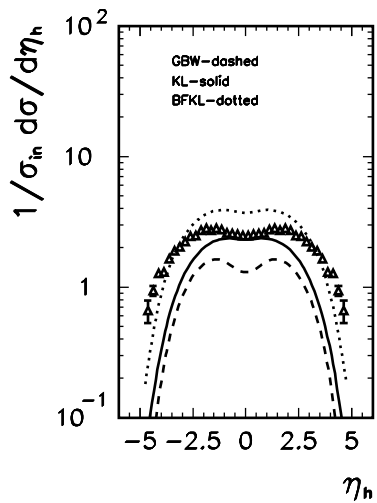}
\end{center}
\caption[*]{
Charged-pion pseudrapidity distribution at W = 200 GeV
for different models of UGD's.
In this calculation $p_{t,h} >$ 0.2 GeV.
The experimental data of the UA5 collaboration are taken
from \cite{UA5_exp}.
\label{fig:eta_glue}
}
\end{center}
\end{figure}


Let us analyze now how the results for pseudorapidity distributions
depend on the choice of the UGD.
In Fig.\ref{fig:eta_glue} I compare pseudorapidity distribution
of charged pions for different models of UGD's.
In this calculation FF from \cite{BKK95} has been used.

In contrast to Ref.\cite{KL01}, where the whole pseudorapidity
distribution, including fragmentation regions, has been well
described in an approach similar to the one presented here,
in the present approach pions produced from the fragmentation
of gluons in the $gg \rightarrow g$ mechanism populate only
midrapidity region,
leaving room for other mechanisms in the fragmentation regions.
These mechanisms involve quark/antiquark degrees of freedom
or leading protons among others. This strongly suggests
that the agreement of the result of the $gg \rightarrow g$
approach with the PHOBOS distributions \cite{PHOBOS} in
Ref.\cite{KL01} in the true fragmentation region is rather due to
approximations made in \cite{KL01} than due to correctness
of the reaction mechanism. In principle, this can be verified
experimentally at RHIC by measuring the $\pi^+ / \pi^-$ ratio
in proton-proton scattering as a function of (pseudo)rapidity
in possibly broad range.
The BRAHMS experiment can do it even with the existing apparatus. 
 
\begin{figure}[!thb]
\begin{center}
  \begin{center}
\includegraphics[width=6cm]{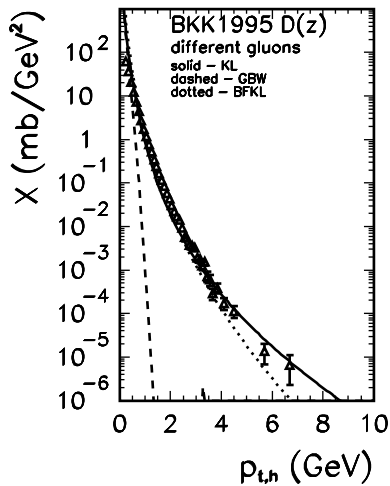}
  \end{center}
\caption[*]{
Transverse momentum distributions of charged pions at W = 200 GeV
for BKK1995 FF and different UGD's.
The experimental data are taken from \cite{UA1_exp}.
\label{fig:pt_glue}
}
\end{center}
\end{figure}

In Fig.\ref{fig:pt_glue} I compare the theoretical transverse
momentum distributions of charged pions obtained with
different gluon distributions with the UA1 collaboration data
\cite{UA1_exp}.
The best agreement is obtained with the
Karzeev-Levin gluon distribution. The distribution with
the GBW model is much too steep in comparison to experimental
data. This is probably due to neglecting QCD evolution
in \cite{GBW_glue}.


\section{Unintegrated parton distributions from the solution
of the CCFM equation}

\begin{figure} 
  \begin{center}
\includegraphics[width=6cm]{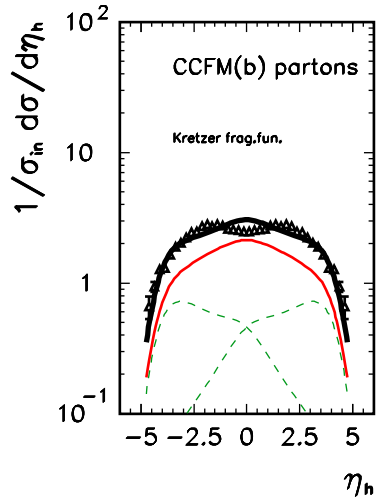}
  \end{center}
\caption[*]{
Pseudorapidity distribution of charged pions at W = 200 GeV
calculated with the CCFM parton distributions.
The experimental data are taken from \cite{UA1_exp}.
The thin solid line is the gluon-gluon contribution while
the dashed lines represent the gluon-(anti)quark
and (anti)quark-gluon contributions. In this calculation
gluon, (anti)quark fragmentation functions from \cite{Kretzer2000}
have been used.
\label{fig:CCFM}
}
\end{figure}

Many unintegrated gluon distributions in the literature
are ad hoc parametrizations of different sets of
experimental data rather than derived from QCD.
Recently Jan Kwieci\'nski and collaborators
\cite{CCFM_b1,GKB03} have shown how to solve so-called CCFM equations
by introducing unintegrated parton distributions in the space
conjugated to the transverse momenta \cite{CCFM_b1}
\begin{equation}
f_q(x,\kappa^2,\mu^2) = \frac{1}{2 \pi}
 \int  \exp \left( i \vec{\kappa} \vec{b} \right) \; 
\tilde f_q(x,b,\mu^2) \; d^2 b \; .
\label{Fourier transform}
\end{equation}
The approach proposed is very convenient to
introduce the nonperturbative effects like
internal (nonperturbative) transverse momentum distributions
of partons in nucleons.
It seems reasonable to include the nonperturbative
effects in the factorizable way
\begin{equation}
\tilde{f}_q(x,b,\mu^2) = 
\tilde{f}_q^{CCFM}(x,b,\mu^2)
 \cdot F_q^{np}(b) \; .
\label{modified_uPDFs}
\end{equation}
In the following, for simplicity, I shall use a flavour and
$x$-independent form factor
\begin{equation}
F_q^{np}(b) = F^{np}(b) = \exp\left(\frac{b^2}{4 b_0^2}\right) \; 
\label{formfactor}
\end{equation}
which describes the nonperturbative effects.
In Fig.\ref{fig:CCFM} I show the distribution in pseudorapidity
of charged pions calculated with the help of the CCFM parton
distributions \cite{GKB03} and the Gaussian form factor
(\ref{formfactor}) with $b_0$ = 0.5 GeV$^{-1}$, adjusted to
roughly describe the UA5 collaboration data.
Now both gluon-gluon and (anti)quark-gluon and gluon-(anti)quark
fussion processes can be included in one consistent framework.
As anticipated the missing up to now terms are more important
in the fragmentation region, although its contribution
in the central rapidity region is not negligible.
More details concerning the calculation will be presented elsewhere
\cite{szczurek2004}.

\section{Conclusions}

I have calculated the inclusive distributions of gluons and
associated charged $\pi$'s in the NN collisions
through the $g g \rightarrow g$ mechanism
in the $k_t$-factorization approach. The results for several
UGD's proposed recently have been compared. The results, especially
$p_{t,h}$ distributions, obtained with different models of
UGD's differ considerably.

Contrary to a recent claim in Ref.\cite{KL01},
I have found that the gluonic mechanism discussed does not describe
the inclusive spectra of charged particles in the fragmentation
region, i.e. in the region of large (pseudo)rapidities for
any UGD from the literature.
I think that at presently available energies the gluonic mechanism
is not the only one.
Some of the missing mechanisms have been estimated
in the approach based on parton distributions
originating from the solution of the CCFM equation.
It was found that the dominant missing mechanisms
are (anti)quark-gluon and gluon-(anti)quark fusion processes.

The existing UGD's lead to the contributions which almost exhaust
the strength at midrapidities and leave room for other mechanisms
in the fragmentation regions. It seems that a measurement at RHIC of
$p_t$ distributions of particles for different rapidities
should be helpful to test unintegrated parton distributions.


\begin{thebibliography}{99}

\bibitem{RHIC}
Proceedings of the Quark Matter 2002 conference, July 2002,
Nantes, France, to be published.

\bibitem{PHOBOS}
B.B. Back et al.(PHOBOS collaboration), Phys. Rev. Lett. {\bf 87}
(2001) 102303-1.

\bibitem{KL01}
D. Kharzeev and E. Levin, Phys. Lett. {\bf B523} (2001) 79.

\bibitem{BRAHMS}
I.G. Bearden et al.(BRAHMS collaboration), Phys. Rev. Lett. {\bf 87}
(2001) 112305.

\bibitem{AKMS94}
A.J. Askew, J. Kwieci\'nski, A.D. Martin and P.J. Sutton,
Phys. Rev. {\bf D49} (1994) 4402.

\bibitem{GBW}
K. Golec-Biernat and M. W\"usthoff, Phys. Rev. {\bf D59} (1999) 014017.

\bibitem{GBW_glue}
K. Golec-Biernat and M. W\"usthoff, Phys. Rev. {\bf D60} (1999)
114023-1.

\bibitem{KMR}
M.A. Kimber, A.D. Martin and M.G. Ryskin, Eur. Phys. J. {\bf C12} 655;\\
M.A. Kimber, A.D. Martin and M.G. Ryskin, Phys. Rev. {\bf D63}
(2001) 114027-1.

\bibitem{Blue95}
J. Bl\"umlein, a talk at the workshop on Deep Inelatic Scattering and
QCD, hep-ph/9506403.


\bibitem{GLR81}
L.V. Gribov, E.M. Levin and M. G. Ryskin, Phys. Lett. {\bf B100}
(1981) 173.

\bibitem{Sz03}
A. Szczurek, Acta Phys. Polon. {\bf 34} (2003) 3191.


\bibitem{BKK95}
J. Binnewies, B.A. Kniehl, G. Kramer, Phys. Rev. {\bf D52} (1995) 4947.

\bibitem{Kretzer2000}
S. Kretzer, Phys. Rev. {\bf D62} (2000) 054001.


\bibitem{UA5_exp}
G.J. Alner et al. (UA5 collaboration), Z. Phys. {\bf C33} (1986) 1.

\bibitem{UA1_exp}
C. Albajar et al. (UA1 collaboration), Nucl. Phys. {\bf B335} (1990) 261.


\bibitem{CCFM_b1}
J. Kwieci\'nski, Acta Phys. Polon. {\bf B33} (2002) 1809.

\bibitem{GKB03}
A. Gawron, J. Kwieci\'nski and W. Broniowski, Phys. Rev. {\bf D68}
(2003) 054001.


\bibitem{szczurek2004}
A. Szczurek, a paper in preparation.


\end{thebibliography}
\end{document}